# Synthesis of high-entropy-alloy-type superconductors (Fe,Co,Ni,Rh,Ir)Zr$_2$ with tunable transition temperature

Md. Riad Kasem[1], Aichi Yamashita[1], Yosuke Goto[1], Tatsuma D. Matsuda[1], Yoshikazu Mizuguchi[1]*

1. Department of Physics, Tokyo Metropolitan University, 1-1, Minami-osawa, Hachioji, 192-0397.

(Corresponding author: mizugu@tmu.ac.jp)

**Abstract**

We report on the synthesis and superconductivity of high-entropy-alloy-type (HEA-type) compounds $Tr$Zr$_2$ ($Tr$ = Fe, Co, Ni, Rh, Ir), in which the $Tr$ site satisfies the criterion of HEA. Polycrystalline samples of HEA-type $Tr$Zr$_2$ with four different compositions at the $Tr$ site were synthesized by arc melting method. The phase purity and crystal structure were examined by Rietveld refinement of X-ray diffraction profile. It has been confirmed that the obtained samples have a CuAl$_2$-type tetragonal structure. From analyses of elemental composition and mixing entropy at the $Tr$ site, the HEA state for the $Tr$ site was confirmed. The physical properties of obtained samples were characterized by electrical resistivity and magnetization measurements. All the samples show bulk superconductivity with various transition temperature ($T_c$). The $T_c$ varied according to the compositions and showed correlations with the lattice constant $c$ and $Tr$-Zr bond lengths. Introduction of an HEA site in $Tr$Zr$_2$ is useful to achieve systematic tuning of $T_c$ with a wide temperature range, which would be a merit for superconductivity application.

# 1. Introduction

Superconductivity is a macroscopic quantum phenomenon typically characterized by zero-resistivity state, Meissner effect, and Josephson effects. Superconductors are used in large-scale application and device applications [1]. Particularly, low-transition temperature (low-$T_c$) metals and alloy, which is mostly Nb or Pb-based alloys, are used for device components like Josephson junctions. Since superconducting devices have been expected to be used in various purposes and scales in near future, development of new superconducting materials, which are easy to use in practical application and have tunable superconducting properties, are needed. In this study, to develop new superconducting materials system having tunability of $T_c$, we focus on high-entropy-alloy-type (HEA-type) compounds.

Recently, superconductivity in HEAs have been getting attention due to the possible robustness of superconducting states under extreme conditions and flexible tuning of superconducting properties by alloying [2-11], since the discovery of a $Ta_{34}Nb_{33}Hf_8Zr_{15}Ti_{11}$ superconductor with $T_c$ = 7.3 K [2]. HEAs are typically defined as alloys composed of five or more elements with a concentration range between 5 to 35at%. Due to extremely high performance as high-temperature materials, structural materials, medical materials, etc., which are most probably resulted from high configurational mixing entropy ($\Delta S_{mix}$) defined as $\Delta S_{mix} = -R \Sigma_i c_i \ln c_i$, where $c_i$ and $R$ are compositional ratio and the gas constant, HEAs have been extensively studied in the field of material science, engineering, chemistry and physics [12,13].

To further extend the field of HEA superconductors, we have synthesized HEA-type compounds, in which the crystal structure possesses an HEA-type site [14-18]. In addition, by Fujita et al., single crystals of $RE$(O,F)BiS$_2$ layered superconductors with an HEA-type rare-earth ($RE$) site were grown [19]. Although the number of the examples of HEA-type superconductors is still limited, various (positive or negative) effect of increased mixing entropy ($\Delta S_{mix}$) to the superconducting characteristics have been observed [14,17,20]. Very recently, we reported the synthesis of $Co_{0.2}Ni_{0.1}Cu_{0.1}Rh_{0.3}Ir_{0.3}Zr_2$ and the emergence of bulk superconductivity with $T_c$ = 8 K [21]. This material has a tetragonal CuAl$_2$-type structure (*I*4/*mcm*, #140) and is a member of $Tr$Zr$_2$ ($Tr$: transition metal). $Tr$Zr$_2$ has been studied as superconducting system with a relatively high $T_c$ up to 11.3 K for $Tr$ = Rh [22-24] and as hydrogen-storage materials [25]. The $T_c$ of those $Tr$Zr$_2$ varies largely by element substitution: $T_c$ = 0.17, 5.5, 1.6, 11.3, 7.5 K for $Tr$ = Fe, Co, Ni, Rh, and Ir, respectively. On the basis of the fact, we considered that the $Tr$Zr$_2$ system is suitable for $T_c$ tuning with a wide temperature range. In this study, we choose $Tr$ = Fe, Ni, Co, Rh, and Ir because pure $Tr$Zr$_2$ compounds with those $Tr$ elements show superconductivity as mentioned above. We have synthesized four HEA-type (Fe,Co,Ni,Rh,Ir)Zr$_2$ polycrystalline samples and observed bulk superconductivity with variable $T_c$ by changing the $Tr$-site compositions.

## 2. Materials and methods

Polycrystalline samples of $Tr$Zr$_2$ with $Tr$ = (A) Fe$_{0.1}$Co$_{0.2}$Ni$_{0.1}$Rh$_{0.3}$Ir$_{0.3}$, (B) Fe$_{0.1}$Co$_{0.3}$Ni$_{0.2}$Rh$_{0.1}$Ir$_{0.3}$, (C) Fe$_{0.2}$Co$_{0.2}$Ni$_{0.2}$Rh$_{0.2}$Ir$_{0.2}$, and (D) Fe$_{0.3}$Co$_{0.2}$Ni$_{0.3}$Rh$_{0.1}$Ir$_{0.1}$ were synthesized by arc melting in Ar atmosphere. Powders of pure Co (99%), Ni (99.9%), Fe (99.9%), Rh (99.9%), and Ir (99.9%), were mixed with a certain composition by mortar and pestle and pelletized into 7mm in diameter. The metal pellet and plates of pure Zr (99.2%) were used as starting materials. To obtain homogeneous sample, arc melting was repeated five times for all the samples. The weight (typically 1 g in total) loss was less than 1 % for all samples. The chemical compositions of the obtained samples were characterized by energy dispersive X-ray fluorescence analysis with a spot size of 0.9 mm on JSX-1000S (JEOL). The phase purity and the crystal structure were examined by powder X-ray diffraction (XRD) with Cu-K$\alpha$ radiation by the $\theta$-$2\theta$ method on Miniflex-600 (RIGAKU) equipped with a high-resolution semiconductor detector D/tex-Ultra. The obtained XRD patterns were refined by the Rietveld method using RIETAN-FP [26], and schematic images of the refined crystal structure were depicted using VESTA [27]. Electrical resistivity was measured by the four-probe method with a current of 5 A on a GM refrigerator system (AXIS). Magnetization measurements were performed by a superconducting quantum interference device (SQUID) on MPMS-3 (Quantum Design) under a magnetic field of 1 mT. The magnetic $T_c$ was estimated as a temperature where magnetization clearly begins to decrease due to the emergence of diamagnetism. Irreversibility temperature ($T_{irr}$), which indicates the emergence of superconducting current, was estimated as the temperature where the zero-field cooling (ZFC) data deviates from field cooling (FC) data. See Supporting Information (Fig. S1) for estimations of $T_c$ and $T_{irr}$.

## 3. Results

The analyzed compositions and estimated $\Delta S_{mix}$ ($Tr$ site) are listed in Table I. Since the actual compositions for the $Tr$ site are close to starting nominal ones, we use starting nominal composition when displaying the results for the samples in this paper. Examined samples are labeled (A)–(D) according to the nominal compositions described in the synthesis part. In addition, the compositional ratio of all the constituent $Tr$-site elements for all the samples satisfy the criterion of HEA, which is 5 to 35 at%, and the estimated mixing entropy $\Delta S_{mix}$ ($Tr$ site) is 1.50$R$, 1.48$R$, 1.61$R$, and 1.52$R$ for sample (A), (B), (C), and (D), respectively.

Figure 1(a) shows the powder XRD patterns for the obtained $Tr$Zr$_2$ samples. Tiny impurity peaks for cubic IrZr$_2$ (space group: No. 227) were observed. The main peaks of the XRD patterns were indexed by the tetragonal CuAl$_2$-type structure (space group: $I4/mcm$). As shown in Fig. 1(b), the XRD peaks shift to higher angles from (A) to (D), which indicates that lattice constants decrease by decreasing average ionic radius of $Tr$. Figure 2 shows the results of the Rietveld refinements for sample (A)–(D). The reliability factor of the Rietveld refinement ($R_{wp}$) in a single-phase analysis was 6.2%, 5.6%, 4.6%, and 3.9% for sample (A), (B), (C), and (D), respectively. Those low $R_{wp}$ support the synthesis of high-purity polycrystalline samples by arc melting and the appropriateness of the CuAl$_2$-type structural model. The structural parameters obtained from the Rietveld analyses are summarized in Table II. Figure 1(c) displays schematic images of the refined crystal structure for sample (C) Fe$_{0.2}$Co$_{0.2}$Ni$_{0.2}$Rh$_{0.2}$Ir$_{0.2}$Zr$_2$, in which the $Tr$ site is occupied by five transition metals. Three Zr-Zr bonds and two $Tr$-Zr bonds were defined as shown in the images.

Figure 3 shows the temperature dependences of electrical resistivity for sample (A)–(D). As shown in Fig. 3(a), zero resistivity was observed at $T_c^{zero}$ = 6.8, 5.7, 4.9, 3.9 K for sample (A), (B), (C), and (D), respectively. As displayed in Fig. 3(b), all the samples show the metallic temperature dependence of resistivity. The large residual resistivity ratio was observed for all the sample, which is a common trend seen in several HEA-type superconductors [2,16-18,21].

Figure 4 shows the temperature dependences of magnetization for sample (A)–(D). All the samples showed large diamagnetic signals. The observed ZFC and FC data show features of a typical type-II superconductor. The ZFC magnetization at 2 K for all the samples exceed a value expected from full shielding volume fraction. Therefore, the superconducting transition observed for all the samples are bulk in nature. Magnetic $T_c^M$ and $T_{irr}$ differs for those four samples: $T_c^M$ = 7.8, 6.7, 5.4, and 4.8 (K) and $T_{irr}$ = 7.0, 5.9, 5.1, and 4.2 K for sample (A), (B), (C), and (D). The $T_c$s estimated from resistivity and magnetization are comparable and show variation according to the composition at the $Tr$ site.

## 4. Discussion

One of the notable trends of this system is that the observed $T_c$s for sample (A)–(D) are close to the *Tr-weighted-average* $T_c$ of pure systems. Since introduction of disorders of the transition metal site sometimes largely affects superconducting properties of transition-metal-based compounds like Fe-based superconductors [26,27], the simple averaging effect on $T_c$ is interesting. Since all the samples show bulk superconductivity, we propose that systematic tuning of $T_c$ can be easily achieved in the $TrZr_2$ system. Such system would be useful for device application containing superconductors. As found in XRD analyses, the lattice constants continuously change according to the ionic radius of elements included in the structure. To obtain information about the relationship between $T_c$ and structural parameters, $T_c^M$ for various $TrZr_2$ are plotted in Figs. 5(a) and 5(b) as functions of lattice constants $a$ and $c$. Those figures contain data points for non-HEA (pure) and related $TrZr_2$.

The lattice constant-$a$ dependence of $T_c$ (Fig. 5(a)) shows monotonous increase in $T_c$ with increasing $a$ for the present HEA-type samples (A)–(D). However, other data points do not show a significant correlation between $T_c$ and $a$. In contrast, the lattice constant $c$ shows a clear correlation with $T_c$ (Fig. 5(b)). Actually, the $Tr$-Zr bonds (Table II) are expanded by increasing concentration of larger $Tr$ such as Rh and Ir, while the Zr-Zr bond length does not show a remarkable dependence on the change in $Tr$-site concentration. Since the electronic bands near the Fermi energy ($E_F$), which is essential for the superconductivity, are contributed by both Zr-$d$ and $Tr$-$d$ electrons [24], the systematic change in bonding states along the $c$-axis, which is mainly composed of $Tr$-Zr bonds, may systematically tunes density of states at the $E_F$ and hence $T_c$ in the $Tr$Zr$_2$ system. We note that the data point for FeZr$_2$ does not obey the relation proposed above. In Ref. 32, the presence of weak ferromagnetism was reported for FeZr$_2$. The presence of magnetism may be related to the low $T_c$ for FeZr$_2$. Although our HEA-type samples contain Fe with various concentration (about 30% for (D)), the presence of Fe does not affect the relation; sample (D) also obeys the trend. The random solution in the HEA-type compounds may be effective to the suppression of magnetic ordering. If the assumption is correct, this fact should positively work on systematic tuning of $T_c$ by compositional manipulation in HEA-type $Tr$Zr$_2$. To obtain further knowledge, synchrotron XRD experiments and/or experiments sensitive to local structure, such as X-ray absorption spectroscopy, are needed for the system.

## 5. Conclusion

We have synthesized polycrystalline samples of new HEA-type $Tr$Zr$_2$ superconductors with $Tr$ = Fe, Co, Ni, Rh, and Ir by arc melting. By Rietveld refinements of the powder XRD patterns, the crystal structures were confirmed to belong to the CuAl$_2$-type tetragonal structure. All the samples show bulk superconductivity with various $T_c$. The $T_c$ shows a correlation with lattice constant $c$ and $Tr$-Zr bond length along the $c$-axis. This can be related to the electronic states contributed by both $Tr$-$d$ and Zr-$d$ electrons near the $E_F$ in $Tr$Zr$_2$. We propose that HEA effects are useful to achieve systematic tuning of $T_c$ with a wide temperature range in $Tr$Zr$_2$, which is possibly due to the systematic tuning of density of states at $E_F$ and suppression of magnetism in high-entropy alloying. Such trend (functionality) would be useful for device application containing superconductors.

**Acknowledgements**

The authors thank O. Miura for their assistance with the experiments. This work was partly supported by JSPS KAKENHI (Grant Number: 18KK0076) and Tokyo Metropolitan Government Advanced Research (Grant Number: H31-1).

**Table I. Actual composition (XRF) and estimated $\Delta S_{mix}$ ($Tr$ site) for the examined $TrZr_2$ samples (A)–(D).**

| Label | Actual $Tr$ composition (XRF) | $\Delta S_{mix}$ ($Tr$ site) |
|---|---|---|
| (A) | $Fe_{0.093}Co_{0.194}Ni_{0.113}Rh_{0.271}Ir_{0.329}$ | $1.50R$ |
| (B) | $Fe_{0.108}Co_{0.297}Ni_{0.202}Rh_{0.073}Ir_{0.320}$ | $1.48R$ |
| (C) | $Fe_{0.190}Co_{0.190}Ni_{0.200}Rh_{0.212}Ir_{0.208}$ | $1.61R$ |
| (D) | $Fe_{0.293}Co_{0.190}Ni_{0.300}Rh_{0.093}Ir_{0.124}$ | $1.52R$ |

**Table II. Structural data and $T_c$ for the $TrZr_2$ samples.**

| Label | (A) | (B) | (C) | (D) |
|---|---|---|---|---|
| Space group | \multicolumn{4}{c}{$I4/mcm$} | | | |
| $a$ (Å) | 6.4711(3) | 6.4671(4) | 6.4517(2) | 6.4406(3) |
| $c$ (Å) | 5.5621(3) | 5.5287(4) | 5.5226(2) | 5.4825(3) |
| $V$ (Å$^3$) | 232.91(2) | 231.23(3) | 229.87(2) | 227.42(2) |
| x (Zr) | 0.1673(3) | 0.1682(3) | 0.1683(2) | 0.1696(2) |
| y (Zr) | 0.6679(3) | 0.6689(3) | 0.6670(2) | 0.6690(2) |
| $R_{wp}$ (%) | 6.2 | 5.6 | 4.6 | 3.9 |
| Zr-Zr 1 (Å) | 3.4056(13) | 3.3994(13) | 3.3990(9) | 3.3827(9) |
| Zr-Zr 2 (Å) | 3.411(3) | 3.4022(13) | 3.404(2) | 3.388(2) |
| Zr-Zr 3 (Å) | 3.164(2) | 3.140(2) | 3.1437(13) | 3.1106(13) |
| Tr-Zr 1 (Å) | 2.779(2) | 2.771(2) | 2.7653(12) | 2.7553(12) |
| Tr-Zr 2 (Å) | 2.784(2) | 2.776(2) | 2.7750(12) | 2.7598(12) |
| $T_c^{zero}$ (K) | 6.8 | 5.7 | 4.9 | 3.9 |
| $T_c^M$ (K) | 7.8 | 6.7 | 5.4 | 4.8 |
| $T_{irr}$ (K) | 7.0 | 5.9 | 5.1 | 4.2 |

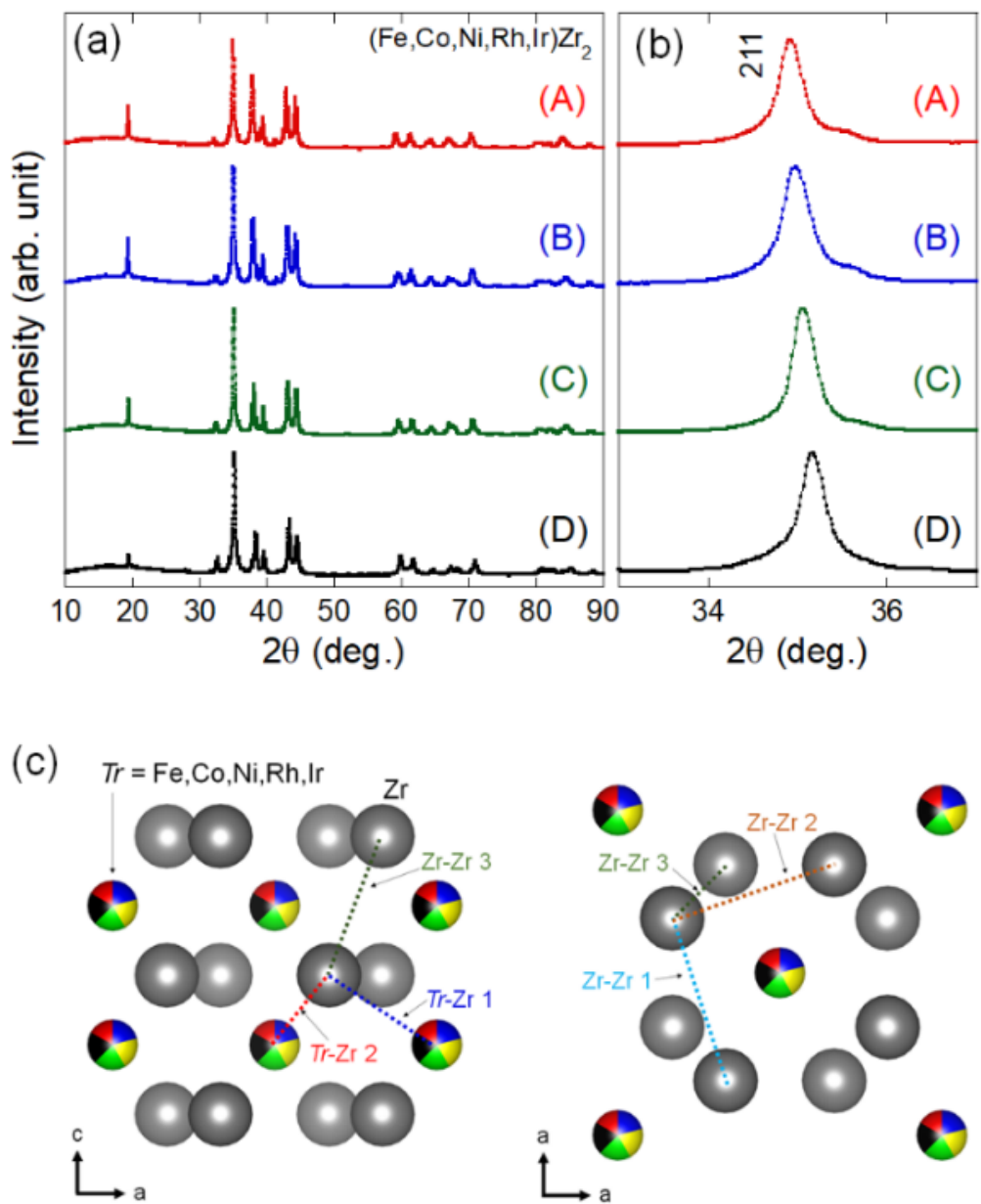

Fig. 1. (a) Powder XRD patterns for the obtained $Tr$Zr$_2$ samples (A)–(D). (b) Zoomed profiles near the 211 peak. (c) Schematic images of the crystal structure of sample (C) Fe$_{0.2}$Co$_{0.2}$Ni$_{0.2}$Rh$_{0.2}$Ir$_{0.2}$Zr$_2$.

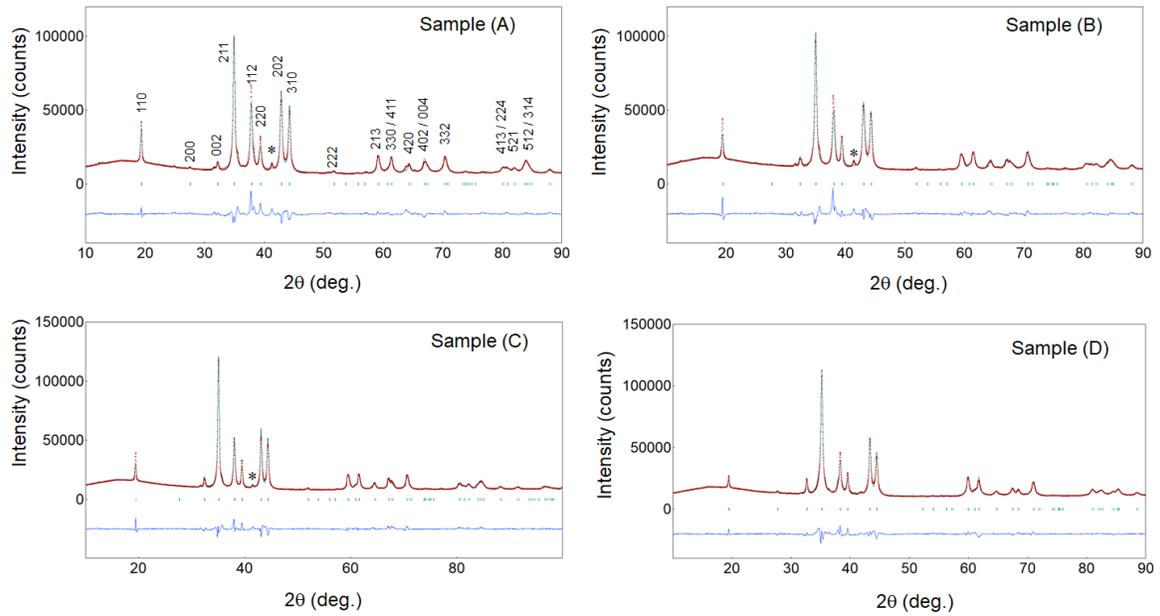

Fig. 2. (a) Powder XRD patterns for the obtained $Tr$Zr$_2$ samples (A)–(D). (b) Zoomed profiles near the 211 peak. (c) Schematic images of the crystal structure of sample (C) Fe$_{0.2}$Co$_{0.2}$Ni$_{0.2}$Rh$_{0.2}$Ir$_{0.2}$Zr$_2$.

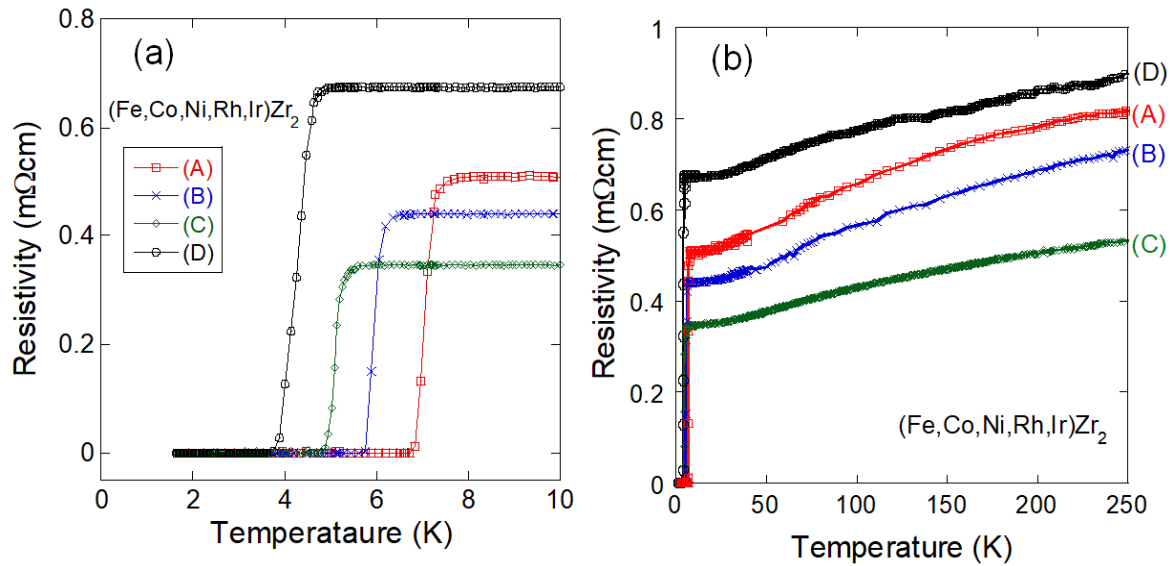

Fig. 3. Temperature dependences of electrical resistivity for $Tr$Zr$_2$ samples (A)–(D) at low temperatures (a) and whole temperature range (b).

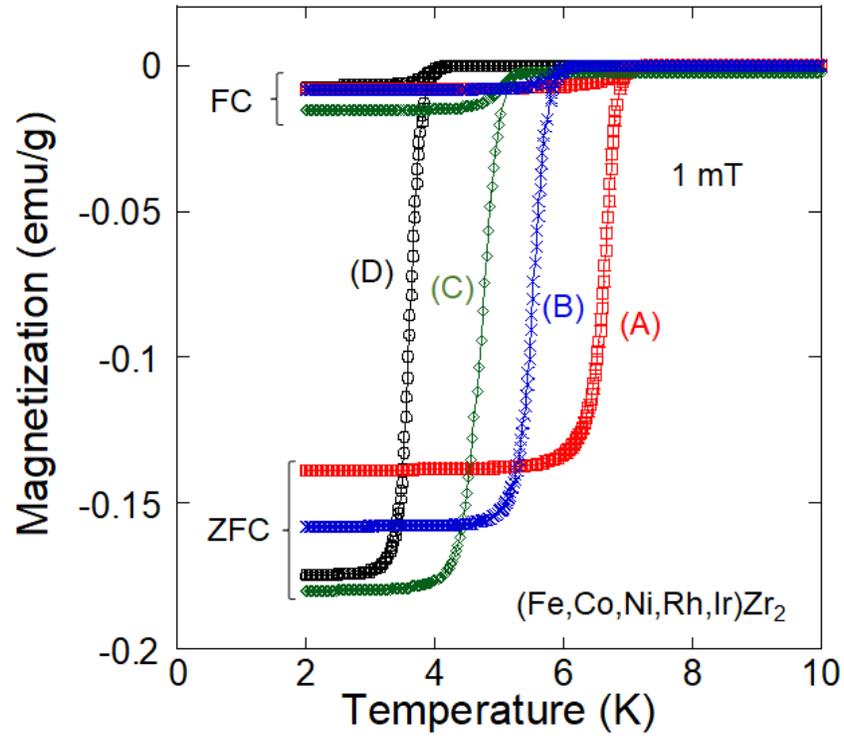

Fig. 4. Temperature dependences of magnetization for *Tr*Zr$_2$ samples (A)–(D).

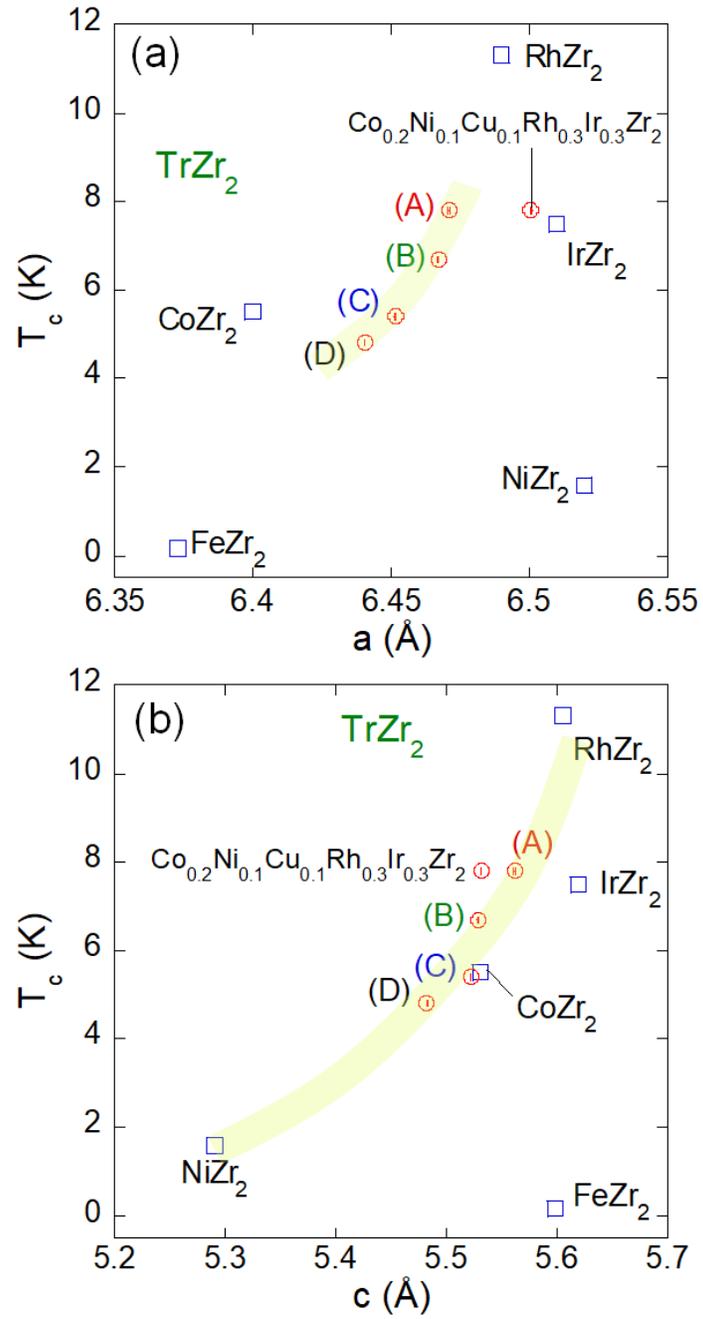

Fig. 5. Lattice constant dependences of $T_c$ for $Tr$Zr$_2$. Data of $T_c$ and lattice constants were taken from Refs. 21, 22, 28-31.

# Synthesis of high-entropy-alloy-type superconductors (Fe,Co,Ni,Rh,Ir)Zr$_2$ with tunable transition temperature

Md. Riad Kasem[1], Aichi Yamashita[1], Yosuke Goto[1], Tatsuma D. Matsuda[1], Yoshikazu Mizuguchi[1]*

2. Department of Physics, Tokyo Metropolitan University, 1-1, Minami-osawa, Hachioji, 192-0397.

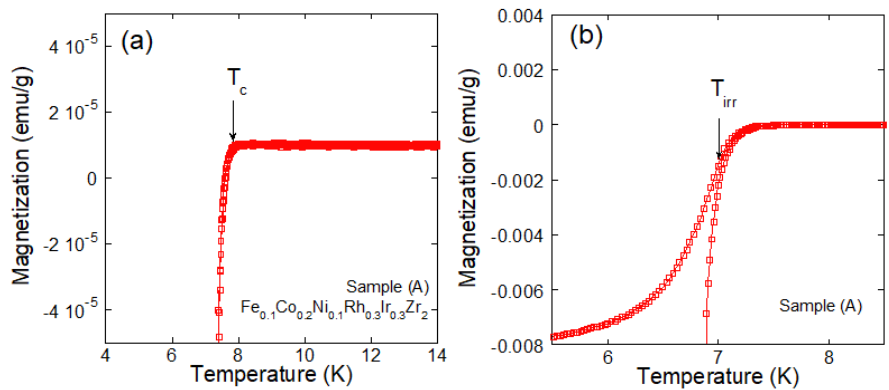

Fig. S1. Estimation for $T_c$ and $T_{irr}$ from the temperature dependence of magnetization for sample (A).